\documentclass[floats,floatfix,showpacs,amssymb,prd,onecolumn,superscriptaddress,nofootinbib,twocolumn]{revtex4-2}

\usepackage{graphicx}
\usepackage{dcolumn}
\usepackage{bm}
\usepackage{amsmath}
\usepackage[colorlinks=true,linkcolor=blue,citecolor=blue,urlcolor=blue]{hyperref}
\usepackage{dsfont}

\usepackage{xcolor,cancel}

\usepackage{xcolor}
\usepackage{ulem}
\usepackage{makecell}

\newcommand\comment[1]{}

\usepackage{romannum}

\begin{document}

\pagenumbering{arabic}

\title{Cosmological evolution with decaying dark matter: an integral-equation approach}

\author{Nanoom Lee}
\email{nanoom.lee@jhu.edu}
\affiliation{William H. Miller III Department of Physics \& Astronomy, Johns Hopkins University, Baltimore, Maryland 21218, USA}

\author{Anna Bencke}
\affiliation{William H. Miller III Department of Physics \& Astronomy, Johns Hopkins University, Baltimore, Maryland 21218, USA}

\author{Marc Kamionkowski}
\affiliation{William H. Miller III Department of Physics \& Astronomy, Johns Hopkins University, Baltimore, Maryland 21218, USA}

\author{Jos\'e Luis Bernal}
\affiliation{Instituto de Física de Cantabria (IFCA), CSIC-Univ. de Cantabria, Avda. de los Castros s/n, E-39005 Santander, Spain}

\begin{abstract}
We present \texttt{CLASSIER-DDM}, an extension of the Boltzmann solver \texttt{CLASSIER} that implements the decaying dark matter (DDM) model via its integral-equation approach. The code handles generic two-body decays of a dark matter particle into two lighter decay products with arbitrary masses, naturally encompassing both massless (dark radiation) and massive (warm) decay products, with their perturbations evaluated via integral equations solved iteratively. We describe the numerical implementation in detail, including the background evolution, the iterative perturbation evolution, and a small-scale analytic approximation.  A modest number of iterations is sufficient to achieve sub-$0.1\%$ convergence in the matter power spectrum and the CMB lensing power spectrum across the observationally relevant parameter space. We find that a hundred momentum bins for the decay product perturbations are sufficient to achieve $\mathcal{O}(0.1\%)$ accuracy in the matter power spectrum today up to $k \sim 3\,{\rm Mpc}^{-1}$. The code achieves $\mathcal{O}(1\,{\rm min})$ runtimes per evaluation, requiring neither a Boltzmann hierarchy nor any fluid approximation, making parameter estimation with the DDM model numerically tractable. We also discuss the impact of DDM on cosmological observables, focusing on the case of two massive decay products. This work further establishes the integral-equation approach as a versatile and efficient framework for modeling non-cold relics in cosmological perturbation theory.
\end{abstract}

\maketitle

\section{Introduction}

One of the chief aims of cosmological surveys such as CMB-S4 \cite{CMB-S4:2016ple, Abazajian:2019eic, Abazajian:2022nyh}, Simons Observatory \cite{SimonsObservatory:2018koc,SimonsObservatory:2019qwx}, CMB-HD \cite{Sehgal:2019ewc,CMB-HD:2022bsz}, DESI \cite{DESI:2016fyo,DESI:2019jxc}, the Rubin Observatory Legacy Survey of Space and Time (LSST) \cite{LSST:2008ijt,LSSTDarkEnergyScience:2012kar, LSSTDarkEnergyScience:2018jkl}, the Nancy Grace Roman Space telescope \cite{Spergel:2015sza}, and Euclid \cite{EUCLID:2011zbd}, is to seek subtle effects in data that may point to nontrivial dark matter properties. One such possibility is a dark matter particle that undergoes decays to warm or hot decay products \cite{Aoyama:2014tga, Wang:2014ina, FrancoAbellan:2020xnr, FrancoAbellan:2021sxk, Montandon:2025xpd, Acharya:2026bmo}. The relevant physics is quite simple: a massive particle at rest decays to two lighter particles that subsequently evolve collisionlessly. However, accurately modeling the effects of such decays on the evolution of cosmological perturbations has traditionally been computationally demanding.

The standard approach for modeling the evolution of non-cold relic perturbations involves solving a Boltzmann hierarchy --- a formally infinite set of coupled differential equations for the Legendre multipole moments of the phase-space distribution --- truncated at some finite maxium multipole. This procedure can be computationally expensive at high precision or on small scales, and truncation-induced numerical artifacts may appear in the perturbation evolution if the hierarchy is not truncated at a sufficiently high multipole \cite{Ma:1995ey, class}. This challenge is particularly acute for the DDM model, where previous work studying decays to one massless and one massive decay product \cite{Aoyama:2014tga, FrancoAbellan:2020xnr, FrancoAbellan:2021sxk} relied on the truncated Boltzmann hierarchy, with Ref.~\cite{FrancoAbellan:2020xnr} further introducing a fluid approximation to reduce the computational cost. 

Recent work has shown that these hierarchies for non-cold relic perturbations can be replaced with a set of a few integral equations \cite{Kamionkowski:2021njk, Ji:2022iji, Lee:2025zym}. This integral-equation approach follows particles along unperturbed geodesics and expresses the perturbations as time integrals over metric sources convolved with spherical Bessel kernels, which can be evaluated efficiently using fast Fourier transforms. This is similar in spirit to the line-of-sight integration method for computing cosmic microwave background (CMB) multipoles \cite{Seljak:1996is}. The approach avoids the numerical artifacts associated with hierarchy truncation, requires no model-specific fluid approximations, and can be faster than the truncated Boltzmann hierarchy, particularly at small scales. It has been implemented for massive neutrinos in \texttt{CLASSIER} (\texttt{CLASS} Integral Equation Revision)\footnote{\href{https://github.com/nanoomlee/CLASSIER}{github.com/nanoomlee/CLASSIER}} \cite{Lee:2025zym}, a modified version of the public Boltzmann solver \texttt{CLASS} \cite{class}, achieving sub-0.01\% accuracy in the matter power spectrum while substantially reducing the computational cost, especially at small scales.

Here we extend this formalism to include decaying dark matter. We present \texttt{CLASSIER-DDM}\footnote{\href{https://github.com/nanoomlee/CLASSIER/tree/CLASSIER-DDM}{github.com/nanoomlee/CLASSIER/tree/CLASSIER-DDM}}, an extension of \texttt{CLASSIER} that handles generic two-body decays of dark matter with arbitrary masses of the decay products, naturally encompassing both massless (dark radiation) and massive (warm) decay products. The integral-equation approach is particularly beneficial in this context since, unlike massive neutrinos, the decay products are produced continuously over time, requiring a fine momentum grid to accurately capture the decay history, which would make the truncated Boltzmann hierarchy approach even more computationally demanding. We describe the numerical implementation in detail, present convergence and performance tests, and discuss the impact of DDM on cosmological observables.\footnote{For cosmological constraints on the DDM model derived using \texttt{CLASSIER-DDM}, we refer the reader to the companion paper \cite{Bencke:2026uws}.} This work further demonstrates the versatility and efficiency of the integral-equation approach as a framework for modeling non-cold relics in cosmological perturbation theory.

The rest of the paper is organized as follows. In Sec.~\ref{sec:background}, we describe the homogeneous evolution of the DDM model, including the evolution of the decaying dark matter density and the phase-space distribution of the decay products. In Sec.~\ref{sec:perturbations}, we present the linear perturbation equations for the decay products and their integral-equation solutions. In Sec.~\ref{sec:implementation}, we describe the numerical implementation of the DDM model in \texttt{CLASSIER-DDM}. In Sec.~\ref{sec:performance}, we assess the numerical accuracy and performance of the code. In Sec.~\ref{sec:results}, we present the evolution of linear perturbations in the presence of DDM and the impact of DDM on cosmological observables. We conclude with a discussion in Sec.~\ref{sec:discussion}. One appendix discusses how to generalize the homogeneous evolution to more than two decay products, another provides details of a small-argument approximation for the convolution integral used in Sec.~\ref{sec:integralsolutions}, and a third compares \texttt{CLASSIER-DDM} with the fluid approximation result of Ref.~\cite{FrancoAbellan:2020xnr}, illustrating the impact of the fluid approximation on the predicted matter power spectrum suppression.

\section{Homogeneous evolution}
\label{sec:background}

We begin by describing the background evolution of the DDM model. We first discuss the evolution of the decaying dark matter density, which we assume to be cold prior to decay, and then derive the homogeneous phase-space distribution of the decay products.

\subsection{Evolution of decaying particle energy density}

We postulate a cold dark matter particle $\chi$ with mass $m_\chi$ that decays with decay rate $\Gamma$.  The comoving number density of this particle is given by
\begin{equation}
    N_\chi(t) = \frac{\rho_\chi(t) a^3(t)}{m_\chi},
\end{equation}
where $t$ is the proper time, $a(t)$ is the scale factor normalized to unity today and $\rho_\chi(t)$ is the physical energy density of particle $\chi$,
\begin{equation}
    \rho_\chi(t) = \rho_\chi^0\, [a(t)]^{-3} e^{-\Gamma t}.
\label{eq:rho_chi}
\end{equation}
Here, $\rho_\chi^0$ is the comoving energy density of decaying dark matter before its decay.

\subsection{Phase-space distribution of decay products}

We now suppose that $\chi$ decays to two particles of mass $m_{\chi_1}$ and $m_{\chi_2}$ (with $m_{\chi_1}+m_{\chi_2} < m_\chi$). The two particles move in opposite directions with momentum
\begin{equation}
     p = \frac12 m_\chi \left[ 1 - 2 (\varepsilon_1^2+\varepsilon_2^2) + (\varepsilon_1^2-\varepsilon_2^2)^2 \right]^{1/2},
     \label{eq:p}
\end{equation}
where $\varepsilon_i \equiv m_{\chi_i}/m_\chi$ is the mass ratio. The dimensionless comoving momentum $q$ of a decay product is its $z=0$ momentum in units of the current CMB temperature $T_0$; i.e., $q=a(\tau_q) p/T_0$, where $\tau_q$ is the conformal time at which it is produced. The energy (in units of the CMB temperature) is $\epsilon_i(q,\tau) = \left[ q^2 +a^2(\tau) m_i^2/T_0^2 \right]^{1/2}$ so that the velocity of a decay product is $q/\epsilon_i(q, \tau)$.\footnote{Note that these definitions of $q$ and $\epsilon$ follow that of Ref.~\cite{Lesgourgues:2011rh}, not Ref.~\cite{Ma:1995ey}. The energy $\epsilon_i$ should not be confused with the mass ratio $\varepsilon_i$.} At production, $a(\tau_q) = qT_0/p$, so all decay products of species i are produced with the same kick velocity, $v_{{\rm kick},i} \equiv  q/\epsilon_i(q, \tau_q) = p/(p^2 + m_{\chi_i}^2)^{1/2}$, independent of $q$, after which their velocity decreases as the momentum redshifts. For equal masses, $v_{{\rm kick},1} = v_{{\rm kick},2} \equiv v_{{\rm kick}} = (1 - \varepsilon^2)^{1/2}$, where $\varepsilon\equiv \varepsilon_1 + \varepsilon_2$ is the total mass retention. Note that the cosmological observables depend only on the mass ratios $\varepsilon_i$ not on the absolute mass $m_\chi$. The phase-space density for the decay products then satisfies,
\begin{eqnarray}
     \frac{\partial f_0(q,\tau)}{\partial\tau} &=& \frac{a(\tau) \Gamma N_\chi(\tau)}{4 \pi q^2} \delta_D\left(q- \frac{a(\tau) p}{T_0} \right) \nonumber \\ 
     &= & \frac{a(\tau) \Gamma N_\chi(\tau)}{4 \pi q^3 {\cal H}(\tau)} \delta_D(\tau- \tau_q) \nonumber\\ 
     &\equiv&\widetilde{f}_0(q)\delta_D(\tau- \tau_q) ,
\label{eqn:uniformevolution}     
\end{eqnarray}
where $\delta_D(x)$ is the Dirac delta function, and ${\cal H}=(da/d\tau)/a$. The solution is then
\begin{eqnarray}
     f_0(q,\tau) &=&\widetilde{f}_0(q) \Theta(\tau-\tau_q),
\label{eq:f0}
\end{eqnarray}
where $\widetilde{f}_0(q)$ is the amplitude of the distribution determined at the time of decay $\tau_q$ and $\Theta(x)$ is the Heaviside step function. The comoving average number density of these decay products is then $N(\tau) = \int\, d^3 q \;f_0(q,\tau)$.\footnote{$f_0$ has units of $({\rm volume)}^{-1}$} Then the homogenous energy density and pressure for decay products are given by
\begin{eqnarray}
\bar{\rho}(\tau) &=& 4\pi a^{-4} T_0 \int q^2 dq\, \epsilon\, f_0(q,\tau), \nonumber \\
\bar{P}(\tau) &=& \frac{4\pi}{3} a^{-4} T_0 \int q^2 dq\, \frac{q^2}{\epsilon}\, f_0(q,\tau).
\label{eq:homogeneous-rho-P}
\end{eqnarray}

The homogeneous evolution of the decaying particles and the two decay products described above can easily be generalized to the case with more than two decay products, as shown in Appendix.~\ref{appendix:f0-general}.

\section{Linear Perturbation Evolution}
\label{sec:perturbations}

We now turn to the evolution of perturbations in the presence of DDM. The phase-space distribution of the decay products for a Fourier mode with wavenumber $k$ can be written as $f(\vec{q},k,\tau) = f_0(q,\tau) + \Delta f(q,k,\mu,\tau)$, where $f_0(q,\tau)$ is the homogeneous background distribution derived in Sec.~\ref{sec:background}. In the synchronous gauge, the perturbation $\Delta f(q,k,\mu,\tau)$ satisfies \cite{Aoyama:2014tga},
\begin{eqnarray}
     \frac{\partial \Delta f}{\partial \tau} + ik\mu\frac{q}{\epsilon}\Delta f + \frac{\partial f_0}{\partial\ln q}\left[ \dot{\eta} - \frac{\dot{h}+6\dot{\eta}}{2}\mu^2\right] =  \frac{\partial f_0}{\partial \tau} \delta_\chi(k,\tau),\nonumber \\
\label{eq:Boltzmann}
\end{eqnarray}
where $\mu$ is the cosine of the angle between ${\bf k}$ and the direction of the particle momentum, $h(k,\tau)$ and $\eta(k,\tau)$ are the synchronous-gauge metric perturbations, dots denote derivatives with respect to conformal time $\tau$, and $\delta_\chi$ is the decaying dark matter density perturbation, which is identical to the cold dark matter (CDM) density perturbation in the synchronous gauge. For brevity, we suppress the $k$ and $\mu$ dependence of $\Delta f$ where the context is clear. This has the solution (for given $q$, $k$, and $\mu$),
\begin{eqnarray}
     \Delta f(q,\tau) &=& \int_0^\tau\, d\tau'\, e^{-i \mu k \chi_q(\tau',\tau)} \left\{   \frac{\partial f_0(q,\tau')}{\partial \tau} \delta_\chi(k,\tau) \right. \nonumber\\
      &+& \left. \frac{ \partial f_0(q,\tau')}{\partial \ln q} \left[- \dot \eta (\tau') +\frac{\dot h(\tau') +6 \dot \eta(\tau')}{2}\mu^2  \right] \right\},~~~~~
\end{eqnarray}
where $\chi_q(\tau_1,\tau_2)\equiv \int_{\tau_1}^{\tau_2} d\tau' q/\epsilon(\tau')$ is the comoving distance traveled by the particle from $\tau_1$ to $\tau_2$.\footnote{Note that $\chi_q$, which follows the notation of Ref.~\cite{Lee:2025zym}, should not be confused with the particle labels $\chi$, $\chi_1$, and $\chi_2$.} The Legendre multipole moments $(\Delta f)_\ell \equiv (i^\ell/2)\int_{-1}^1 \Delta f(\mu) P_\ell(\mu)$, then become
\begin{eqnarray}
     (\Delta f)_l(q,\tau) &=&   \int_0^{\tau} d\tau'\; \Bigg\{  j_\ell[k \chi_q(\tau',\tau)] \frac{\partial f_0(q, \tau')}{\partial \tau} \delta_\chi(\tau') \nonumber \\
      &&+ \frac{\partial f_0(q, \tau')}{\partial \ln q} \Bigg[   -j_\ell[k\chi_q(\tau',\tau)] \dot{\eta}  \nonumber \\
      &&\qquad\qquad\qquad - j''_\ell[k\chi_q(\tau',\tau)] \frac{\dot{h} + 6\dot{\eta}}{2} \Bigg]\Bigg\},~~~~~
    \label{eq:Psi_ell}
\end{eqnarray}
where $j''_\ell(x) \equiv d^2 j_\ell(x) / dx^2$. 

For two-body decays, the background distribution in Eq.~\eqref{eq:f0} satisfies
\begin{equation}
\frac{\partial f_0(q,\tau)}{\partial \ln q}
=-\frac{\widetilde{f}_0(q)}{\mathcal{H}(\tau_q)} \delta(\tau-\tau_q) + \frac{d\widetilde{f}_0(q)}{d\ln q} \Theta(\tau-\tau_q)\,.
\end{equation}
Substituting this expression into Eq.~\eqref{eq:Psi_ell} yields
\begin{eqnarray}
(\Delta f)_\ell(q,\tau) = (\Delta f)^{\rm init}_\ell(q,\tau)+(\Delta f)^{\rm cont}_\ell(q,\tau),
\label{eq:Df_ell}
\end{eqnarray}
where $(\Delta f)^{\rm init}_\ell(q,\tau)$ represents the contribution sourced at the initial time (e.g., the decay time $\tau_q$) and subsequently propagated to later times through free streaming,
\begin{eqnarray}
  &&   (\Delta f)^{\rm init}_\ell(q,\tau) \equiv  \widetilde{f}_0(q) \delta_\chi(\tau_q) j_\ell[k\chi_q(\tau_q,\tau)] \nonumber\\
  &&  \quad\qquad\qquad\qquad + \frac{\widetilde{f}_0(q)}{\mathcal{H}(\tau_q)} \Bigg\{\dot{\eta}(\tau_q)j_\ell[k\chi_q(\tau_q,\tau)] \nonumber\\
  &&  \quad\qquad\qquad\qquad\qquad\qquad + \frac{\dot{h}(\tau_q) + 6\dot{\eta}(\tau_q)}{2}j''_\ell[k\chi_q(\tau_q,\tau)]\Bigg\}.\nonumber\\
  \label{eq:Df_ell-init}
\end{eqnarray}
Note that in the limit $\tau \rightarrow \tau_q$, the term $(\Delta f)^{\rm init}_\ell(\tau, q)$ reproduces exactly the initial perturbation derived in Ref.~\cite{Aoyama:2014tga} (see their Appendix~A). For $\tau > \tau_q$, this contribution simply corresponds to the free-streamed evolution of that initial condition.

The remaining integral contribution is
\begin{eqnarray}
(\Delta f)^{\rm cont}_\ell(q,\tau)
&\equiv& \frac{d\widetilde{f}_0}{d\ln q}
\int_{\tau_q}^{\tau} d\tau' \Bigg\{ - j_\ell\!\left[k\chi_q(\tau',\tau)\right]\dot{\eta}(\tau') \nonumber\\ 
&& \qquad\quad- j''_\ell\!\left[k\chi_q(\tau',\tau)\right] \frac{\dot{h}(\tau')+6\dot{\eta}(\tau')}{2} \Bigg\},\nonumber\\
\label{eq:Df-cont}
\end{eqnarray}
which is continuously sourced at $\tau>\tau_q$. Introducing the variable $\xi(q,\tau,k) \equiv k \chi_q(\tau_q,\tau)$, the integral becomes a convolution \cite{Kamionkowski:2021njk,Ji:2022iji,Lee:2025zym}
\begin{eqnarray}
    (\Delta f)^{\rm cont}_\ell(q,\tau) &=& \frac{d \widetilde{f}_0}{d \ln q} \int_{0}^{\xi}  \frac{\epsilon(\xi')}{kq}\, d\xi' \, \Big\{ 
    -j_\ell(\xi-\xi') \dot{\eta} (\xi') \nonumber\\
    && \quad\qquad\qquad - j''_\ell(\xi-\xi') \frac{\dot{h}(\xi') + 6\dot{\eta}(\xi')}{2} \Big\}.\nonumber\\
    \label{eq:I-convolution}
\end{eqnarray}
This integral contribution together with the contribution sourced at the decay time [Eq.~\eqref{eq:Df_ell-init}] determines the total perturbations of multipole moments [Eq.~\eqref{eq:Df_ell}]. Given the phase-space density perturbation multipoles, the contributions from the decay products to the stress-energy tensor are given by
\begin{eqnarray}
\delta \rho (\tau) &=& 4\pi a^{-4} T_0 \int q^2 dq\, \epsilon\, (\Delta f)_0(q,\tau), \nonumber \\
\delta P(\tau) &=& \frac{4\pi}{3} a^{-4} T_0 \int q^2 dq\, \frac{q^2}{\epsilon}\, (\Delta f)_0(q,\tau), \nonumber \\
(\bar{\rho} + \bar{P})\theta(\tau)
&=& 4\pi k a^{-4} T_0 \int q^2 dq\, q\, (\Delta f)_1(q,\tau), \nonumber \\
(\bar{\rho} + \bar{P})\sigma(\tau)
&=& \frac{8\pi}{3} a^{-4} T_0 \int q^2 dq\, \frac{q^2}{\epsilon}\, (\Delta f)_2(q,\tau),~~
\label{eq:decay-perturb}
\end{eqnarray}
where $P$, $\theta$, $\sigma$ are the pressure, velocity divergence, and anisotropic stress. Note that the factor of $T_0$ arises because the phase-space density $f$ has units of (volume)$^{-1}$, by definition [Eq.~\eqref{eq:f0}]. Since only the $\ell=0$, 1, and 2 multipole moments enter the stress-energy tensor, we evaluate the integral equations only for these moments. Contributions from all higher multipoles are implicitly captured as the full perturbation system reaches self-consistency through the iterative scheme.

\section{Numerical Implementation}
\label{sec:implementation}

We implement the decaying dark matter model into \texttt{CLASSIER} \cite{Lee:2025zym}, a modified version of \texttt{CLASS} \cite{class} in which the integral-equation method has been implemented for massive neutrinos, resulting in a new code \texttt{CLASSIER-DDM}. The implementation consists of two parts: the background evolution, described in Sec.~\ref{sec:bg}, and the perturbation evolution, described in Sec.~\ref{sec:pert}. Throughout, we work in the synchronous gauge.

\subsection{Background evolution}
\label{sec:bg}

The background evolution of the DDM model is handled within \texttt{CLASSIER-DDM}'s background module. The decaying dark matter density evolves as $\rho_\chi = \rho_\chi^0 a^{-3} e^{-\Gamma t(a)}$ [Eq.~\eqref{eq:rho_chi}], where we define  $\rho_\chi^0 \equiv f \rho_{\rm cdm,0}$ with the unstable fraction $f$  and the CDM energy density today $\rho_{\rm cdm,0} = 3(H_0/h)^2 \omega_{\rm cdm} / (8\pi G)$.\footnote{This is the total CDM energy density today in the absence of decays.} We solve the standard background evolution equations together with the homogeneous phase-space distribution of the decay products, $f_0(q, \tau)$.

The decay products' phase-space distribution $f_0(q, \tau)$ is discretized on a grid of momentum bins $q_i \equiv a(\tau_{q_i}) p/T_0$, where $\tau_{q_i}$ is the conformal time at which the decay product corresponding to bin $q_i$ is produced. Each bin $q_i$ thus corresponds to decay products produced at the scale factor
\begin{equation}
    a_{q_i} = \frac{q_i \, T_0}{p},
\end{equation}
where $p$ is the physical momentum of the decay products at production, Eq.~\eqref{eq:p}. While evolving the standard background equations, each $q$-bin is filled when $a = a_{q_i}$ with the amplitude $\widetilde{f}_0(q_i)$ [Eq.~\eqref{eqn:uniformevolution}] and thereafter remains constant. At each time step, the energy density and pressure of the decay products are evaluated from the filled phase-space distribution [Eq.~\eqref{eq:homogeneous-rho-P}]. To make this approach numerically straightforward, we use a forward Euler integrator instead of \texttt{CLASS}'s default adaptive integrator for the background evolution. By default, we use 1,000 bins for the background momentum grid,\footnote{The corresponding parameter in \texttt{CLASSIER-DDM} is \texttt{N\_q\_ddm\_bg}.} evenly spaced in $\ln q$ between $q_{\rm min} \equiv 10^{-4}(p/T_0)$ and $q_{\rm max} = {\rm min}(10\,a_{\rm decay}, 1)\,(p/T_0)$.

Due to the discreteness of the momentum grid, when computing the background energy density and pressure at a given scale factor $a_{q_i} < a < a_{q_{i+1}}$, there is a missing contribution from decay products with momenta $q_i < q < q_{\rm dec}(a) \equiv ap/T_0$, since these intermediate momenta are not represented in the $q$-grid. To correct for this, we add a contribution from the interval $[q_i, q_{\rm dec}(a)]$, evaluated at the current $a$ and $H$. Once $a$ reaches $a_{q_{i+1}}$, the bin $q_{i+1}$ is filled and its contribution naturally replaces the correction, ensuring a smooth and consistent transition. This ensures that the contribution from the most recently produced decay products is not missed due to the finite bin spacing.

A separate, coarser momentum grid is defined for the perturbation evolution, as fewer momentum bins are sufficient compared to the background evolution since the decay product perturbations are not solved simultaneously with other perturbations. This grid is filled during the background evolution, during which we also compute and store $\chi_q(\tau_q, \tau) \equiv \int_{\tau_q}^{\tau} d\tau' q/\epsilon(\tau')$ for each decay product since its production, which is needed for the integral-equation method described in Sec.~\ref{sec:pert}. This separate perturbation momentum grid does not contribute to background energy densities or pressures.

\subsection{Perturbation evolution}
\label{sec:pert}

For the perturbation evolution, we take an iterative approach. In the 0th iteration, since the decay product perturbations are not yet available, we approximate the total density perturbation of the DDM sector, comprising the decaying dark matter and its decay products, as
\begin{equation}
    \delta\rho_\text{DDM}^{(0)} = 
    \big[\rho_\chi + \left(\rho_{\chi_1}-3P_{\chi_1}\right)  + \left(\rho_{\chi_2}-3P_{\chi_2}\right)
    \big]\delta_c,
\label{eq:0th-iter}
\end{equation}
where $\delta_c$ is the CDM density perturbation. The combination $\rho_{\chi_i} - 3P_{\chi_i} = \int \frac{d^3q}{(2\pi)^3} f_0(q) \,m_i^2/\epsilon_i$ represents the rest-mass energy density of each decay product, which naturally accounts for the time-dependent transition between relativistic and non-relativistic regimes: it vanishes when the decay products are highly relativistic ($P_{\chi_i} \approx \rho_{\chi_i}/3$), and reduces to $\rho_{\chi_i}$ in the non-relativistic limit ($P_{\chi_i} \ll \rho_{\chi_i}$).\footnote{The convergence of the iterative scheme could in principle be further improved by incorporating a rough analytic modeling of the free-streaming suppression of decay product perturbations already at the 0th iteration, for example by multiplying the decay product contributions in Eq.~\eqref{eq:0th-iter} by a damping factor of the form $e^{-(k\lambda_{\rm fs})^2}$, where $\lambda_{\rm fs}$ is the comoving free-streaming length of the decay products. We leave such an improvement to future work.} We then use the perturbation variables $\dot{h}(\tau)$, $\dot{\eta}(\tau)$, and $\delta_\chi(\tau)[=\delta_c(\tau)]$ to evaluate the phase-space perturbations of the decay products through the integral-equation method, and evolve the rest of perturbation system with these provided. This is our 1st iteration, and we iterate this process for a pre-determined number of iterations for good convergence.

The integral-equation method for the DDM decay products is implemented on top of the existing massive neutrino implementation in \texttt{CLASSIER} \cite{Lee:2025zym}. However, for the calculations shown in this work we apply the integral-equation method only to the DDM decay products, while retaining the standard truncated Boltzmann hierarchy with the fluid approximation of \texttt{CLASS} \cite{class} for massive neutrinos, as the DDM signal dominates over the massive neutrino effects, hence high precision for the massive neutrino perturbations is unnecessary.

\subsubsection{Integral solutions of perturbed Boltzmann equation}
\label{sec:integralsolutions}

Two key aspects arise in the implementation of the integral-equation method for the DDM model. First, since each momentum $q$ has a different decay time $\tau_q$, the onset of perturbations differs for each momentum, making the lower limit of the convolution integral momentum-dependent and requiring careful numerical treatment. Second, since the convolution integral starts at the decay time $\tau_q$, and we focus on decays occurring after recombination ($z_d \lesssim 1100$, where $z_d$ is the characteristic decay redshift), the integration range $[\tau_q, \tau_0]$ is limited to late times. This eliminates the need for the super-/near-horizon and sub-horizon split employed in Ref.~\cite{Lee:2025zym}. Apart from these, the implementation closely follows Ref.~\cite{Lee:2025zym}.

The total multipole moments $(\Delta f)_\ell$ consist of two contributions [Eq.~\eqref{eq:Df_ell}]: the term sourced at the decay time $\tau_q$, $(\Delta f)_\ell^{\rm init}$ [Eq.~\eqref{eq:Df_ell-init}], which is already analytic and requires no special numerical treatment, and the continuously-sourced term $(\Delta f)_\ell^{\rm cont}$ [Eq.~\eqref{eq:I-convolution}], which requires dedicated numerical evaluation. The main implementation lies in how the latter is evaluated. For the perturbation evolution, we define, by default, 100 momentum bins evenly spaced in $q$ between $q_{\rm min} \equiv 10^{-4}p/T_0$ and $q_{\rm max}={\rm min}(10\,a_{\rm decay}, 1)\,p/T_0$.\footnote{The corresponding parameter in \texttt{CLASSIER-DDM} is \texttt{N\_q\_ddm}. Note that in our implementation, there are two possible choices for binning scheme: \texttt{ddm\_qbin\_scheme\,=\,linear} or \texttt{hybrid}. The linear scheme defines an evenly spaced grid in $q$, while the hybrid scheme combines two grids: one evenly spaced in $q$ and one evenly spaced in $\ln q$, each covering half of the momentum bins. A time-dependent weight is assigned to each grid, smoothly transitioning from the $\ln q$-spaced grid at early times to the $q$-spaced grid at late times. This hybrid scheme is designed to better capture both the early-time impact of DDM, which affects the primary CMB spectra, and the late-time evolution, where a linear $q$-spacing provides better resolution. We therefore recommend the hybrid scheme for decays whose decay epoch overlaps with recombination. Throughout this paper, we use the linear scheme, as we focus on $P_m(k)$ and $C_L^{\phi\phi}$, which are primarily sensitive to the late-time evolution.}

As described in Sec.~\ref{sec:perturbations}, the continuously-sourced contribution to the multipole moments, $(\Delta f)_\ell^{\rm cont}$ [Eq.~\eqref{eq:I-convolution}], can be rewritten as a convolution,
\begin{eqnarray}
    (\Delta f)^{\rm cont}_\ell(q,\tau)  &\equiv&\int_{-\infty}^\infty d\xi'\; G(\xi') K(\xi-\xi'),
\end{eqnarray}
where
\begin{eqnarray}
G(\xi) &\equiv&
\begin{cases}
-\dot{\eta}\frac{\epsilon}{kq}\frac{d \widetilde{f}_0}{d\ln q} & {\rm for} \quad 0 \leq\xi\leq\xi_0, \\
0 & {\rm otherwise},
\end{cases} \\
\quad K(\xi) &\equiv&
\begin{cases}
j_\ell(\xi) & {\rm for} \quad 0 \leq\xi\leq\xi_0, \\
0 & {\rm otherwise},
\end{cases}
\label{eq:GA}
\end{eqnarray}
or, for the second term,
\begin{eqnarray}
G(\xi) &\equiv&
\begin{cases}
 -\frac{\dot{h}+6\dot{\eta}}{2}\frac{\epsilon}{kq}\frac{d \widetilde{f}_0}{d\ln q} & {\rm for} \quad 0 \leq\xi\leq\xi_0, \\
 0 & {\rm otherwise},
 \end{cases}\\
 K(\xi) &\equiv&
 \begin{cases}
  j''_\ell(\xi) & {\rm for} \quad 0 \leq\xi\leq\xi_0, \\
  0 & {\rm otherwise},
\end{cases}
\label{eq:GB}
\end{eqnarray}
where $\xi_0 \equiv \xi(q, \tau_0,k)$ and $\tau_0$ is the conformal time today. These convolutions are evaluated in Fourier space using NUFFTs.

Following Ref.~\cite{Lee:2025zym}, we decompose the source function as
\begin{equation}
G(\xi) = G_0(\xi) + \Delta G(\xi),
\end{equation}
where $G_0(\xi) \equiv a + b\,\xi + c\,\xi^2 + d\,\xi^3$ is defined so that $\Delta G(\xi)$ and its first derivative vanish at the endpoints $\xi=0$ and $\xi=\xi_0$. The convolution with $G_0(\xi)$ is performed analytically (see Table~II of Ref.~\cite{Lee:2025zym}), while the convolution with $\Delta G(\xi)$ is evaluated using non-uniform FFTs via \texttt{FINUFFT} (Flatiron Institute Nonuniform Fast Fourier Transform) \cite{Barnett:2019qbm,Barnett:2020nufft_aliasing}, with Fourier coefficients of $K(\xi)$ computed analytically (see Table~I of Ref.~\cite{Lee:2025zym}). That is, we evaluate
\begin{equation}
     I(\xi) = \int_{-\infty}^\infty\, d\xi'\,\Delta G(\xi') K(\xi-\xi').
\end{equation}
We first evaluate the Fourier coefficients of $\Delta G(\xi)$ by approximating the integral as a sum,
\begin{eqnarray}
    \widehat{\Delta G}(\omega_l)
    &\approx&  \sum_{j=1}^{N^{(\xi)}} w_j^{\rm GL} \Delta G(\xi_j^{\rm GL}) e^{i \omega_l \xi_j^{\rm GL}},
\end{eqnarray}
at $N^{(\omega)} \equiv 10N^{(\xi)}+1$ uniform frequencies with $\Delta \omega_l= \pi/(10\xi_0)$ centered at $\omega_l=0$, where $\xi_j^{\rm GL}$ are the Gauss-Legendre abscissas spaced in $0\leq\xi\leq\xi_0$ and $w_j^{\rm GL}$ are the corresponding weights. By default we set $N^{(\xi)}$ to be $k$-dependent:\footnote{This default setting is optimized for $k_{\rm max} \sim 3\,{\rm Mpc}^{-1}$; for applications requiring higher $k_{\rm max}$, larger values of $N^{(\xi)}$ may be needed to maintain the same level of accuracy.}
\begin{equation}
N^{(\xi)} \equiv \begin{cases}1000 &{\rm if\;} k\leq0.1\;{\rm Mpc}^{-1},\\
5000 &{\rm otherwise.}
\end{cases}
\label{eq:N2-xi}
\end{equation}
We then multiply by the Fourier coefficients of the kernel $\hat{K}(\omega_l)$ and perform the inverse Fourier transform back to configuration space:
\begin{eqnarray}
     I(\xi) 
       &\approx& \sum_{l=1}^{N^{(\omega)}} \Delta \omega_l \,e^{-i \xi \omega_l} \hat K(\omega_l) \widehat{\Delta G}(\omega_l).
\label{eqn:I1ofxi}       
\end{eqnarray}
These evaluations are performed only for the $\ell=0$, 1, and 2 multipole moments, which enter the Einstein equations.

In Appendix~\ref{appendix:approx-direct-conv}, we provide an alternative treatment of the convolution integral for small $\xi$, where the FFT method can be inaccurate as it does not converge smoothly to zero as $\xi \rightarrow 0$. This direct convolution approach is used for $\xi < 1$ regardless of $\xi_0$. Furthermore, when $\xi_0 < 1$, the direct convolution replaces the NUFFT evaluation entirely, which is both more accurate and more efficient in this regime.

\subsubsection{Small-scale analytic approximation}

In this subsection, we develop the analytic small-scale approximation for the integral contribution in Eq.~\eqref{eq:I-convolution} following Ref.~\cite{Lee:2025vgv} in which such an approximation is developed for massive neutrinos.

The integral solution in Eq.~\eqref{eq:Df_ell} consists of two parts: (i) $(\Delta f)^{\rm init}_\ell(\tau,q)$ [Eq.~\eqref{eq:Df_ell-init}], sourced at $\tau=\tau_q$ and then free-streamed, and (ii) the continuously-sourced term $(\Delta f)^{\rm cont}_\ell(\tau,q)$ [Eq.~\eqref{eq:Df-cont}]. The former is already analytic, as it is determined entirely by the values of the sources ($\dot{\eta}$, $\dot{h}$, $\delta_\chi$) at the decay time and propagated forward. The latter term can be interpreted as the particular solution of the following differential equation,
\begin{eqnarray}
\frac{\partial\Delta f}{\partial\tau} + ik\mu\frac{q}{\epsilon}\Delta f + \frac{\partial\widetilde{f}_0}{\partial\ln q}
\left[ \dot{\eta} - \frac{\dot{h}+6\dot{\eta}}{2}\mu^2 \right] =0,
\label{eq:Boltzmann-approx}
\end{eqnarray}
which is identical in form to the collisionless Boltzmann equation solved in Ref.~\cite{Lee:2025vgv}, except that the evolution now begins at $\tau=\tau_q$. Therefore, the small-scale analytic approximation of Ref.~\cite{Lee:2025vgv} can be applied directly by replacing the prefactor $d\ln f_0/d\ln q$ with $d\widetilde{f}_0/d\ln q$. Following Ref.~\cite{Lee:2025vgv}, we define the source term $S(\xi)$ after dropping the $\dot{\eta}$ term,
\begin{equation}
S(\xi) \equiv \frac{\epsilon}{kq}\frac{d \widetilde{f}_0}{d\ln q} \frac{\dot{h}+6\dot{\eta}}{2}.
\end{equation}
The small-scale approximation is then\footnote{Note that there is no homogeneous part in the approximation as $(\Delta f)_\ell^{\rm cont}(\tau_q,q)=0$.}
\begin{eqnarray}
&&(\Delta f)_0^{\rm cont, approx}(\xi) = (\Delta f)_{{\rm QS},0}(\xi) \nonumber\\
&&\qquad\qquad\quad - \sum_{\ell} (-1)^\ell (2 \ell + 1) j_{\ell}(\xi) (\Delta f)_{{\rm QS},\ell}(\xi_*), \label{eq:Psi0-ss} \\
&&(\Delta f)_1^{\rm cont, approx}(\xi) = (\Delta f)_{{\rm QS},1}(\xi) \nonumber\\
&&\qquad\qquad\quad+ \sum_{\ell} (-1)^{\ell}(2 \ell + 1) j_{\ell}'(\xi) (\Delta f)_{{\rm QS},\ell}(\xi_*), ~~~~~~\label{eq:Psi1-ss}\\
&&(\Delta f)_2^{\rm cont, approx}(\xi) = (\Delta f)_{{\rm QS},2}(\xi) \nonumber\\
&&\qquad\qquad - \sum_{\ell} (-1)^{\ell}(2 \ell + 1) \frac{3 j_{\ell}''(\xi) + j_\ell(\xi)}{2} (\Delta f)_{{\rm QS},\ell}(\xi_*),\nonumber\\\label{eq:Psi2-ss}
\end{eqnarray}
with
\begin{eqnarray}
(\Delta f)_{{\rm QS},0}(\xi) &=& S'(\xi),\\
(\Delta f)_{{\rm QS},1}(\xi) &=& \frac 13 S(\xi),\\
(\Delta f)_{{\rm QS},2}(\xi) &=& 0,\label{eq:Delta_l}
\end{eqnarray}
where primes denote derivatives with respect to $\xi$. Combining with Eq.~\eqref{eq:Df_ell-init}, the final small-scale approximation is
\begin{equation}
(\Delta f)^{\rm approx}_\ell(\tau,q)  =  (\Delta f)^{\rm init}_\ell(\tau,q) + (\Delta f)^{\rm cont, approx}_\ell(\tau,q).
\label{eq:Df_ell-approx}
\end{equation}
We use either Eq.~\eqref{eq:Df_ell} or Eq.~\eqref{eq:Df_ell-approx} depending on whether the following criterion for the small-scale approximation is satisfied:
\begin{equation}
s(k, q, \tau_0) \left(\frac{\Gamma}{H_0}\right) < 0.025,
\label{eq:criterion}
\end{equation}
where $s(k, q, \tau_0) \equiv H_0 \epsilon(q, \tau_0) / kq$ is the ratio between the comoving distance free-streamed in a Hubble time at the present time $\tau_0$ and the perturbation comoving wavelength. The additional factor $\Gamma/H_0$ accounts for the fact that a larger decay rate implies an earlier onset of the approximation, at which point the modes may not yet be well inside the horizon and the quasi-static assumption is less justified. The threshold value on the right-hand side is determined empirically, and we find that this criterion yields $\lesssim 0.1\%$ accuracy, speeding up the code by up to a factor of 2. Together, the iterative integral-equation scheme and the small-scale analytic approximation form the core of \texttt{CLASSIER-DDM}, enabling accurate and efficient evaluation of DDM perturbations across a wide range of model parameters.

\section{Numerical Accuracy and Performance}
\label{sec:performance}

\begin{figure*}[ht!]
\centering
\includegraphics[width = 0.49\linewidth,trim= 00 10 00 0]{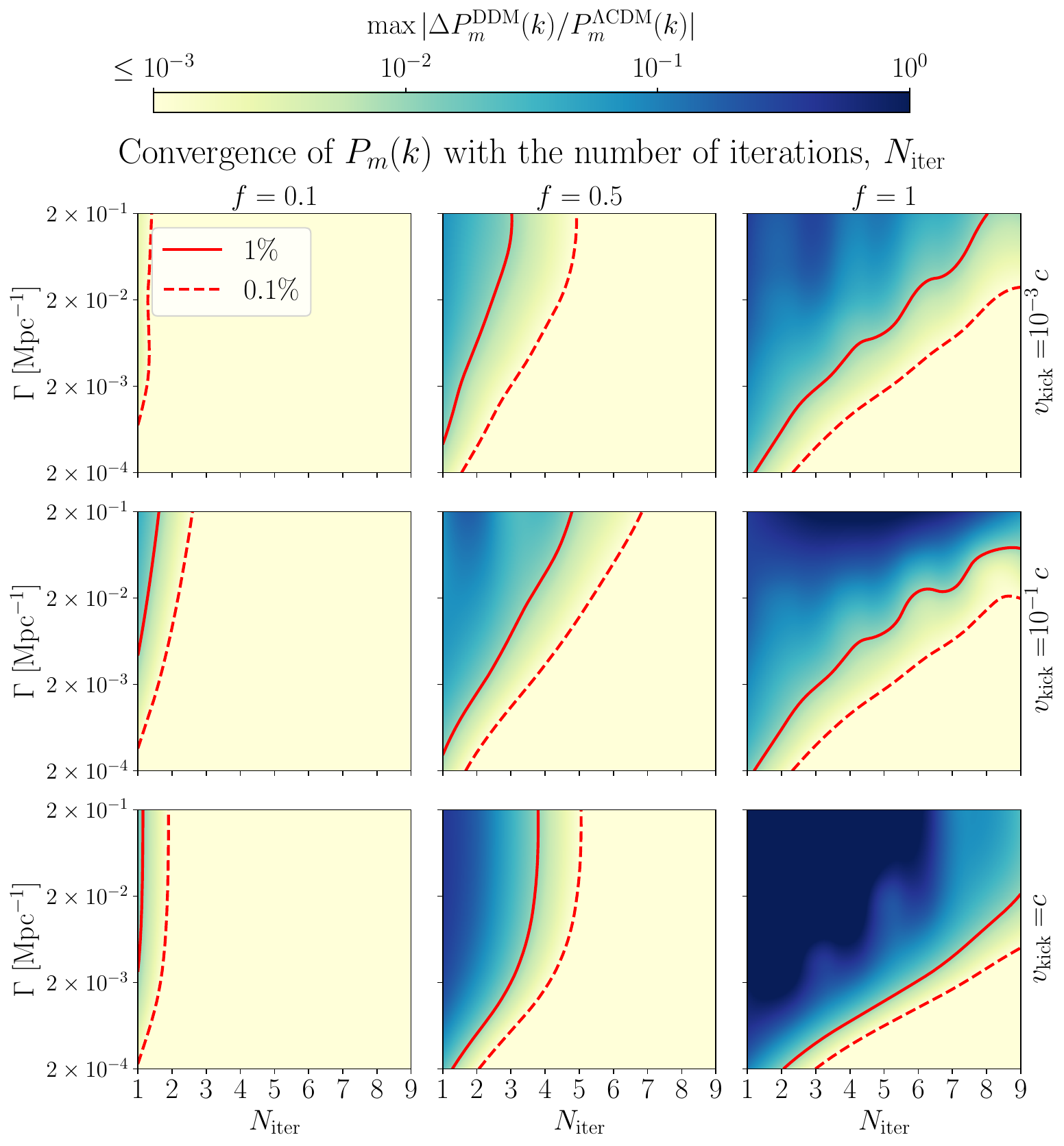}
\includegraphics[width = 0.49\linewidth,trim= 00 10 00 0]{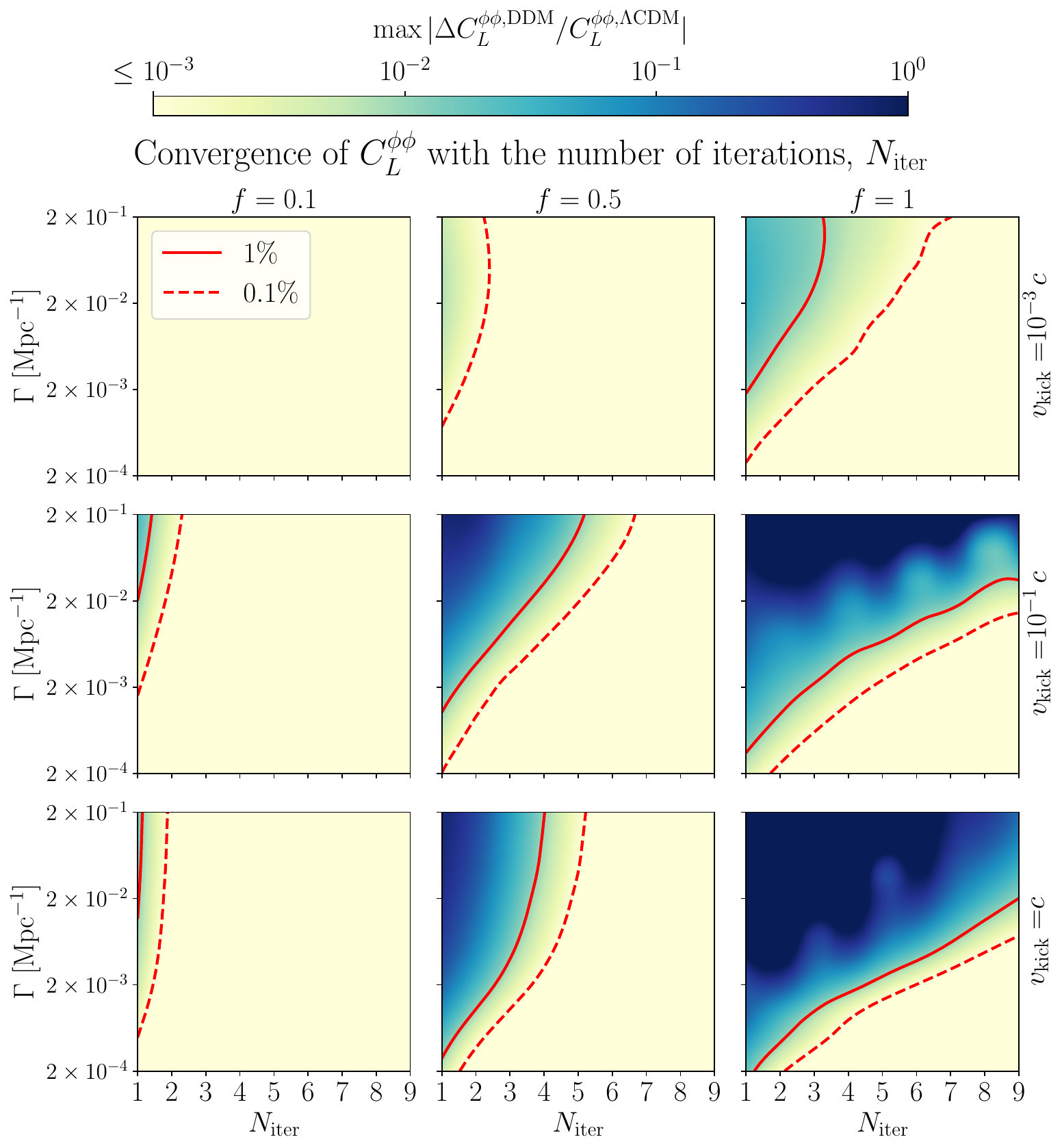}
\caption{Convergence of the matter power spectrum today $P_m(k)$ (left) and the CMB lensing power spectrum $C_L^{\phi\phi}$ (right) with the number of iterations $N_{\rm iter}$, for three values of the decaying fraction $f = 0.1,\, 0.5,\, 1$ (columns) and kick velocity $v_{\rm kick} = 10^{-3}c,\, 10^{-1}c,\, c$ (rows). The color shows the maximum fractional difference between the result at $N_{\rm iter}$ and the reference at $N_{\rm iter} = 10$, normalized by the $\Lambda$CDM power spectrum: ${\rm max}\,|\Delta P_m^{\rm DDM}(k)/P_m^{\Lambda{\rm CDM}}(k)|$ evaluated up to $k_{\rm max} \simeq 3\,{\rm Mpc}^{-1}$ (left), and ${\rm max}\,|\Delta C_L^{\phi\phi,\rm DDM}/C_L^{\phi\phi,\Lambda{\rm CDM}}|$ up to $L=3000$ (right). The solid and dashed red contours mark the 1\% and 0.1\% thresholds, respectively. Convergence improves with increasing $N_{\rm iter}$ and degrades with increasing $f$ and $\Gamma$, though remains good even for $f \to 1$ at small $\Gamma$. The lensing spectrum $C_L^{\phi\phi}$ converges faster than $P_m(k)$ across all parameter combinations shown, as it is an integrated quantity along the line of sight that is less sensitive to the small-scale power suppression driving the slower convergence in $P_m(k)$.}
\label{fig:converg-Pk}
\end{figure*}

We assess the numerical accuracy and performance of \texttt{CLASSIER-DDM} by examining the convergence of the iterative integral-equation method and the computational cost of the code. For simplicity, we assume equal masses for the two decay products throughout this section, in which case their perturbations are identical and need only be computed once. We have verified that the convergence is qualitatively unchanged for unequal masses, while the runtime roughly doubles, as the integral equations must then be evaluated for each decay product separately. We have checked that 100 momentum bins for the phase-space distribution perturbations are sufficient to achieve $\mathcal{O}(0.1\%)$ accuracy in the matter power spectrum today up to $k \sim 3\,{\rm Mpc}^{-1}$. Throughout this paper, we therefore set \texttt{N\_q\_ddm}$=100$ as the default value unless explicitly stated otherwise. For the convergence tests presented in this section, however, we use \texttt{N\_q\_ddm}$=200$ to obtain smoother results; the qualitative conclusions are unchanged.

\subsection{Convergence}
\label{sec:convergence}

We assess the convergence of the iterative integral-equation method by comparing the matter power spectrum $P_m(k)$ and the CMB lensing power spectrum $C_L^{\phi\phi}$ computed with $N_{\rm iter}$ iterations against a reference computed with $N_{\rm iter}=10$ iterations, across the DDM parameter space $(\Gamma, f, v_{\rm kick})$. As a convergence metric, we use the maximum fractional difference between the result at $N_{\rm iter}$ and the reference, normalized by the $\Lambda$CDM power spectrum: ${\rm max}\,|\Delta P_m^{\rm DDM}(k)/P_m^{\Lambda{\rm CDM}}(k)|$ evaluated up to $k_{\rm max} \simeq 3\,{\rm Mpc}^{-1}$, and ${\rm max}\,|\Delta C_L^{\phi\phi,\rm DDM}/C_L^{\phi\phi,\Lambda{\rm CDM}}|$ up to $L=3000$. This choice of normalization avoids spuriously large fractional differences at scales where $P_m^{\rm DDM}(k)$ is strongly suppressed. Furthermore, since large deviations from $\Lambda$CDM are strongly disfavored by current observations, this criterion is practically well-motivated for the parameter space of interest. Here we focus on $P_m(k)$ and $C_L^{\phi\phi}$ as they are the most sensitive probes of the free-streaming of the decay products. The convergence of other primary CMB spectra such as $C_\ell^{TT}$, $C_\ell^{TE}$, and $C_\ell^{EE}$ is expected to be faster, as these are less sensitive to the small-scale power suppression induced by the decays occurring after recombination.

Figure~\ref{fig:converg-Pk} shows these quantities as a function of $N_{\rm iter}$ and $\Gamma$, for three representative values of the decaying fraction $f = 0.1,\, 0.5,\, 1$ (columns) and kick velocity $v_{\rm kick} = 10^{-3}c,\, 10^{-1}c,\, c$ (rows). Several trends are apparent. First, convergence improves monotonically with $N_{\rm iter}$ across the entire parameter space, as expected. Second, $C_L^{\phi\phi}$ converges faster than $P_m(k)$ across all parameter combinations, as it is an integrated quantity along the line of sight that is less sensitive to the small-scale power suppression driving the slower convergence in $P_m(k)$.

The dependence on $v_{\rm kick}$ is non-monotonic: convergence is slightly slower for $v_{\rm kick}=10^{-1}c$ than for $v_{\rm kick}=c$. This may be understood physically as follows. In the ultra-relativistic limit ($v_{\rm kick} \to c$), the decay products free-stream very efficiently and $\rho_{\chi_i} - 3P_{\chi_i} \to 0$, so the 0th-iteration approximation [Eq.~\eqref{eq:0th-iter}] correctly suppresses the DDM contribution to $\delta\rho_{\rm DDM}^{(0)}$, providing a good starting point. At intermediate velocities ($v_{\rm kick} \sim 10^{-1}c$), however, the decay products are semi-relativistic, and the 0th-iteration approximation is neither fully suppressed nor fully accurate, leading to slightly slower convergence. While Fig.~\ref{fig:converg-Pk} shows the convergence tests for equal masses, we find that the convergence is qualitatively the same for unequal decay-product masses.

For small decaying fractions ($f \lesssim 0.1$), $N_{\rm iter} = 2$ is sufficient to achieve $\mathcal{O}(0.1\%)$ convergence across all $v_{\rm kick}$ and $\Gamma$ values considered. For larger decaying fractions ($f \gtrsim 0.5$) or faster decay rates ($\Gamma \gtrsim 10^{-2}\;{\rm Mpc}^{-1}$), more iterations are needed, and for the most extreme parameter combinations the convergence can be slow. We note, however, that for small decay rates ($\Gamma \lesssim 10^{-3}\;{\rm Mpc}^{-1}$), the convergence remains good even for large decaying fractions $f \to 1$, as the DDM sector contributes little to the perturbation evolution at any given time. The slow convergence is therefore specific to the combination of large $f$ and large $\Gamma$. Such extreme parameter combinations imply a substantial deviation from $\Lambda$CDM, which provides an excellent fit to current data, and are unlikely to be of observational interest.

\subsection{Runtime}
\label{sec:runtime}

Table~\ref{tab:runtime} shows the runtimes of \texttt{CLASSIER-DDM} for different numbers of iterations $N_{\rm iter}$ and kick velocities $v_{\rm kick}$, with $k_{\rm max} \sim 3\,{\rm Mpc}^{-1}$ and with a hundred momentum bins (\texttt{N\_q\_ddm}$=100$). The runtime scales roughly linearly with $N_{\rm iter}$, as each iteration requires a similar amount of computation. The dependence on $v_{\rm kick}$ reflects two approximations employed by the code: at small $v_{\rm kick}$, more momentum bins are handled by the small-$\xi$ direct convolution approximation described in Appendix~\ref{appendix:approx-direct-conv}, while at large $v_{\rm kick}$, more bins satisfy the validity criterion of the small-scale analytic approximation of Sec.~\ref{sec:pert}. Both approximations reduce the computational cost relative to the full NUFFT convolution, making the intermediate case $v_{\rm kick}=0.01c$ the most expensive. The runtime is not sensitive to the other DDM parameters $f$ and $\Gamma$ in a direcy way, although note that higher values of $f$ and $\Gamma$ require more iterations to converge (see Fig.~\ref{fig:converg-Pk}).

\begin{table}[!ht]
  \centering
  \begin{tabular}{c|c|c|c}
  $N_{\rm iter}$ & ~$v_{\rm kick}=0.001\,c$~ & ~~$v_{\rm kick}=0.01\,c$~~ & ~~~$v_{\rm kick}=0.1\,c$~~~ \\
  \hline\hline
  $1$  & 60~sec  & 85~sec  & 75~sec \\
  \hline
  $2$  & 100~sec  & 145~sec   & 125~sec \\
  \hline
  $3$ & 140~sec & 205~sec  & 175~sec \\
  \end{tabular}
  \caption{Runtimes on four threads (16-inch MacBook Pro--M4 chip, Nov 2024 release) with $\sim600$ $k$-modes up to $k_{\rm max}\sim 3\,{\rm Mpc}^{-1}$, for different numbers of iterations $N_{\rm iter}$ and kick velocities $v_{\rm kick}$. Here we assume equal masses for the two decay products, in which case their perturbations are identical and need only to be computed once.}
  \label{tab:runtime}
\end{table}

We emphasize that these runtimes are achieved without truncating the Boltzmann hierarchy or employing any fluid approximation for the decay product perturbations. The integral-equation method provides a numerically accurate treatment of the collisionless Boltzmann equation, equivalent to solving the full Boltzmann hierarchy without truncation, hence free from the numerical artifacts associated with hierarchy truncation. Unlike fluid approximations, it requires no model-specific calibrations and is applicable to arbitrary decay product phase-space distributions.

We estimate the efficiency gain of the integral-equation approach over the truncated Boltzmann hierarchy using the code of Ref.~\cite{FrancoAbellan:2020xnr}, which implements DDM with one massless and one massive decay product, as a reference. We find that our code, which can handle two massive decay products, can be more than an order of magnitude faster than the truncated Boltzmann hierarchy approach, under the precision settings required to achieve $\sim 0.1\%$ accuracy up to $k_{\rm max} \sim 3\,{\rm Mpc}^{-1}$. The exact factor for the improvement depends on the DDM parameters, too. This advantage becomes even more significant at smaller scales: the runtime of the truncated Boltzmann hierarchy scales roughly as $k_{\rm max}$ due to the rapidly oscillating perturbations at large $k$ at fixed maximum multipole $\ell_{\rm max}$\footnote{In principle, to maintain the same level of accuracy at larger $k_{\rm max}$, $\ell_{\rm max}$ must be increased accordingly \cite{Ma:1995ey, class}, in which case the runtime of the truncated Boltzmann hierarchy scales as $k_{\rm max}^2$ making the advantage of \texttt{CLASSIER-DDM} even more pronounced.}, while that of \texttt{CLASSIER-DDM} is expected to scale more mildly, approximately as $k_{\rm max}^{1/2}$ \cite{Lee:2025zym}. Furthermore, Appendix~\ref{appendix:comparison} demonstrates that the fluid approximation employed in Ref.~\cite{FrancoAbellan:2020xnr} introduces errors at the percent level in the predicted power spectrum suppression and does not capture the oscillatory features in $P_m(k)$, both of which are accurately reproduced by \texttt{CLASSIER-DDM}.

Altogether, achieving $\mathcal{O}(1\,{\rm min})$ runtimes per evaluation makes parameter estimation with the DDM model numerically tractable, opening the door to comprehensive explorations of the DDM parameter space with current and future cosmological surveys.

\section{Results}
\label{sec:results}

In this section, we present our numerical results from \texttt{CLASSIER-DDM} implementation. We first discuss the evolution of linear perturbations in the presence of DDM in Sec.~\ref{sec:pertevo}, and then present the impact of DDM on the matter power spectrum $P_m(k)$ and the CMB lensing power spectrum $C_L^{\phi\phi}$ in Sec.~\ref{sec:observables}, which are the cosmological observables most sensitive to the free-streaming of the decay products. Throughout, we assume a fiducial cosmology consistent with the Planck 2018 $\Lambda$CDM best-fit parameters \cite{Planck2018}, with one massive neutrino species of mass $m_\nu=0.06\,{\rm eV}$, and equal masses for the two decay products, in which case the kick velocity is $v_{\rm kick} = (1-\varepsilon^2)^{1/2}$, where $\varepsilon \equiv (m_{\chi_1} + m_{\chi_2})/m_\chi$ is the total mass retention.

\subsection{Evolution of linear perturbations with DDM}
\label{sec:pertevo}

\begin{figure}[ht!]
\centering
\includegraphics[width = \linewidth,trim= 00 10 00 0]{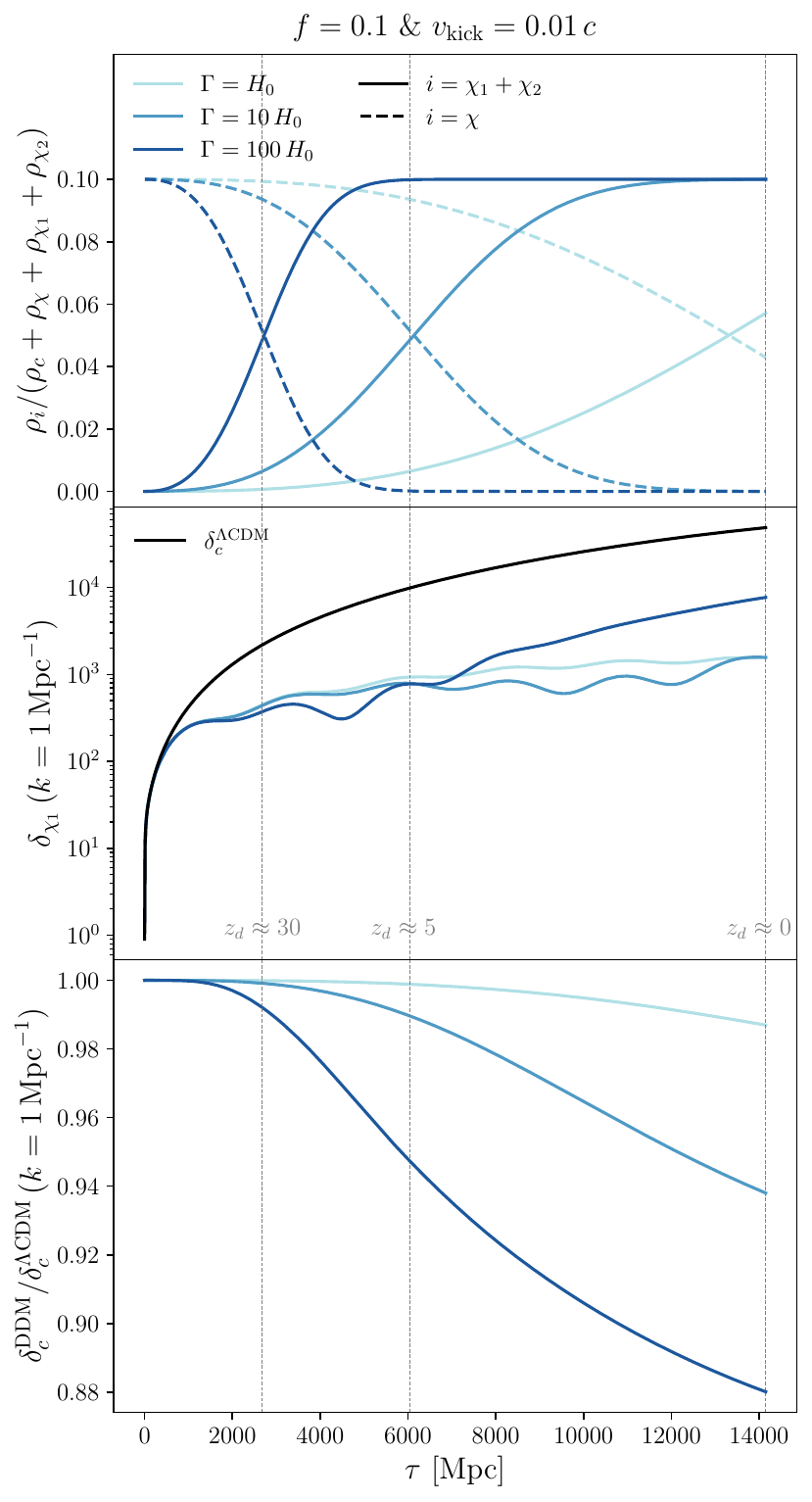}
\caption{Evolution of perturbations in the DDM model for fixed $f=0.1$ and $v_{\rm kick}=0.01c$, varying the decay rate $\Gamma = H_0, 10H_0, 100H_0$ (corresponding to $z_d \approx 0, 5, 30$). Top: background energy densities of the decaying dark matter $\chi$ and the combined decay products $\chi_1+\chi_2$ as a fraction of the total energy density. Middle: decay product density perturbation $\delta_{\chi_1}(k=1\,{\rm Mpc}^{-1})$, compared with the $\Lambda$CDM CDM perturbation $\delta_c^{\Lambda{\rm CDM}}$ (black). Note that the presented perturbations are computed with 2,000 momentum bins ($\texttt{N\_q\_ddm=2000}$), and smoothed for better presentation. Bottom: ratio of the CDM density perturbation in the DDM model to that in $\Lambda$CDM. The ratio falls below $1-f=0.9$, indicating that the suppression of CDM growth exceeds the naive expectation from mass removal alone, due to the gravitational feedback of the free-streaming decay products.}
\vspace{-0.2in}
\label{fig:pertevo_Gamma}
\end{figure}

\begin{figure}[ht!]
\centering
\includegraphics[width = \linewidth,trim= 00 10 00 0]{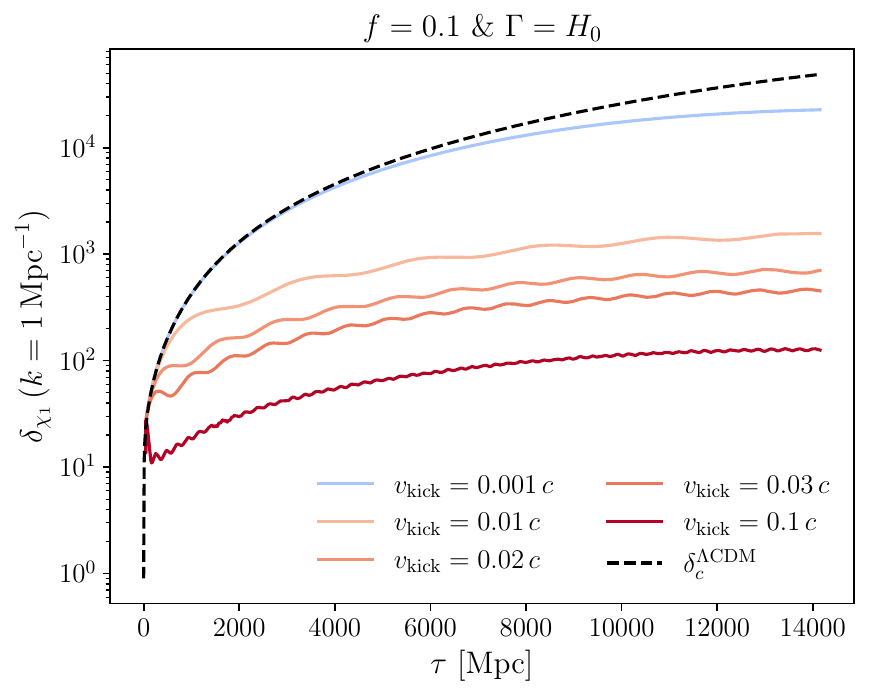}
\caption{Decay product density perturbation $\delta_{\chi_1}(k=1\,{\rm Mpc}^{-1})$ for fixed $f=0.1$ and $\Gamma=H_0$, varying $v_{\rm kick}$. The black dashed line shows $\delta_c^{\Lambda{\rm CDM}}$ for reference. In the limit $v_{\rm kick} \to 0$, the decay product perturbation approaches $\delta_c^{\Lambda{\rm CDM}}$, recovering the CDM behavior. As $v_{\rm kick}$ increases, the overall amplitude of $\delta_{\chi_1}$ is suppressed and the oscillations become more rapid, reflecting the faster accumulation of phase differences among decay products with different momenta.}
\label{fig:pertevo_vkick}
\end{figure}

Figure~\ref{fig:pertevo_Gamma} shows, for fixed $f=0.1$ and $v_{\rm kick}=0.01c$ and three values of the decay rate $\Gamma = H_0, 10H_0, 100H_0$ (corresponding to characteristic decay redshifts $z_d \approx 0, 5, 30$ respectively), the background energy densities (top), the decay product density perturbation $\delta_{\chi_1}(k=1\,{\rm Mpc}^{-1})$ (middle), and the ratio of the CDM density perturbation in the DDM model to that in $\Lambda$CDM, $\delta_c^{\rm DDM}/\delta_c^{\Lambda{\rm CDM}}$ (bottom).

The top panel illustrates the background evolution. The decaying dark matter $\chi$ transfers its energy to the decay products $\chi_1+\chi_2$, with larger $\Gamma$ resulting in an earlier and more rapid transfer. The middle panel shows that the density perturbations of decay product $\delta_{\chi_1}$ begins to grow after the decay time $\tau_q$ and subsequently develops oscillatory features at late times. These oscillations arise from interference between contributions from decay products with different momenta $q$, each of which is produced at a different time $\tau_q$ and free-streams a different comoving distance $\chi_q(\tau_q, \tau)$, accumulating a different phase $k\chi_q(\tau_q,\tau)$ by the time $\tau$. The superposition of these contributions with different phases leads to interference. The characteristic period of these oscillations scales as $1/(kv_{\rm kick})$, reflecting the fact that $v_{\rm kick}$ sets the typical free-streaming velocity and hence the rate at which phase differences accumulate.

The bottom panel shows the ratio $\delta_c^{\rm DDM}/\delta_c^{\Lambda{\rm CDM}}$, which quantifies the suppression of CDM perturbation growth induced by DDM. Notably, this ratio falls below $1-f=0.9$, which would be the naive expectation if DDM simply reduced the CDM density by a fraction $f$ with no further effect. The additional suppression arises because the free-streaming decay products escape from gravitational potential wells, reducing the overall growth rate of the remaining CDM perturbations. This gravitational feedback amplifies the power deficit beyond the simple mass-removal effect, and is more pronounced for larger $\Gamma$ since earlier decays allow more time for this effect to accumulate.

Figure~\ref{fig:pertevo_vkick} shows $\delta_{\chi_1}(k=1\,{\rm Mpc}^{-1})$ for fixed $f=0.1$ and $\Gamma=H_0$, varying $v_{\rm kick}$. In the limit $v_{\rm kick} \to 0$, the decay product perturbation approaches $\delta_c^{\Lambda{\rm CDM}}$, recovering the CDM behavior as expected. As $v_{\rm kick}$ increases, the overall amplitude of $\delta_{\chi_1}$ is suppressed, as the faster free-streaming decay products escape from gravitational potential wells more efficiently. In addition, larger kick velocities lead to more rapid oscillations with a shorter period, since a larger $v_{\rm kick}$ causes phase differences between momentum bins to accumulate more rapidly.

\subsection{Impact on cosmological observables}
\label{sec:observables}

We now present the impact of DDM on the matter power spectrum $P_m(k)$ and the CMB lensing power spectrum $C_L^{\phi\phi}$, which are the observables most sensitive to the free-streaming of the decay products.

\begin{figure}[ht!]
\centering
\includegraphics[width = \linewidth,trim= 00 10 00 0]{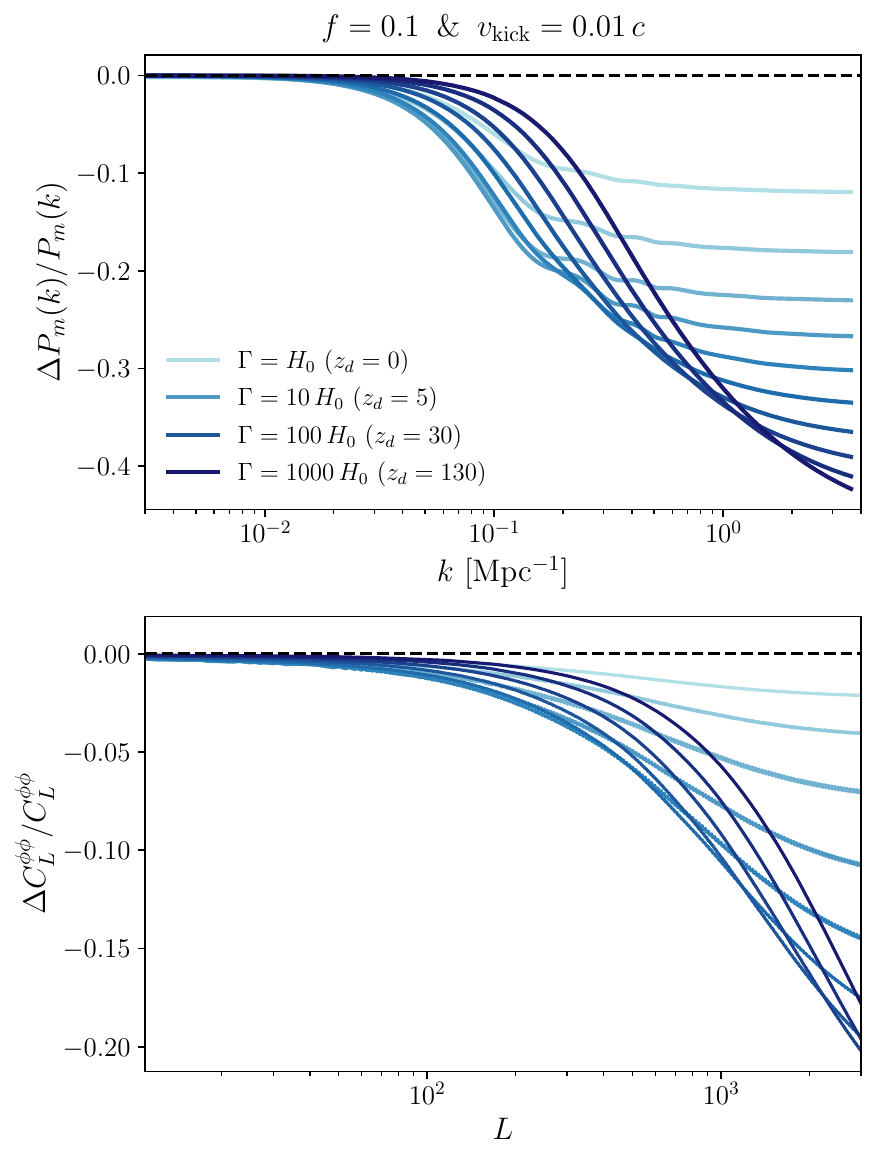}
\caption{Fractional deviations from $\Lambda$CDM in the matter power spectrum $\Delta P_m(k)/P_m(k)$ (top) and the CMB lensing power spectrum $\Delta C_L^{\phi\phi}/C_L^{\phi\phi}$ (bottom), for fixed $f=0.1$ and $v_{\rm kick}=0.01c$, varying the decay rate $\Gamma$. Increasing $\Gamma$ shifts the onset of suppression to larger $k$ (smaller scales), consistent with the scaling $\lambda_{\rm fs}\propto(1+z_d)^{-1/2}$ \cite{Bencke:2026uws}. The suppression in $C_L^{\phi\phi}$ directly reflects the range of $k$ at which $P_m(k)$ is suppressed.}
\label{fig:vary_Gamma}
\end{figure}

\begin{figure}[ht!]
\centering
\includegraphics[width = \linewidth,trim= 00 10 00 0]{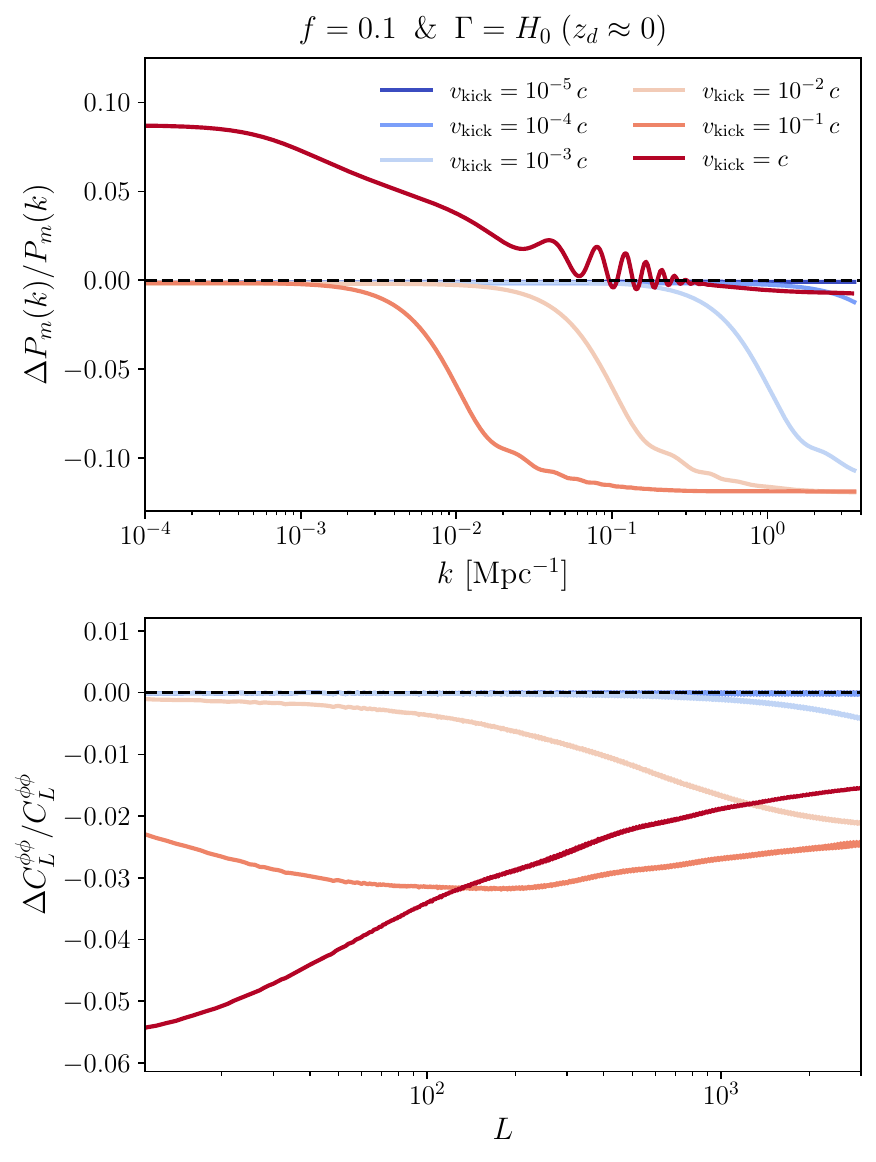}
\caption{Same as Fig.~\ref{fig:vary_Gamma}, but varying $v_{\rm kick}$ at fixed $f=0.1$ and $\Gamma=H_0 \approx 2\times 10^{-4}\;{\rm Mpc}^{-1}$. Increasing $v_{\rm kick}$ shifts the onset of suppression to smaller $k$ (larger scales), reflecting the larger comoving free-streaming length $\lambda_{\rm fs}\propto v_{\rm kick}$. At $v_{\rm kick}=10^{-5}c$ the suppression is negligible, consistent with the $\Lambda$CDM limit. At large $v_{\rm kick} \to c$, the decay products are highly relativistic and the decay effectively converts matter into radiation, modifying the expansion history and leading to a slight enhancement of $P(k)$ on large scales \cite{Aoyama:2014tga}.}
\label{fig:vary_vkick}
\end{figure}

Figures~\ref{fig:vary_Gamma} and \ref{fig:vary_vkick} show the fractional deviations from $\Lambda$CDM, $\Delta P_m(k)/P_m(k)$ and $\Delta C_L^{\phi\phi}/C_L^{\phi\phi}$, for fixed $f=0.1$ while varying either $\Gamma$ or $v_{\rm kick}$, respectively. In both cases, the deviations are negative, reflecting the suppression of structure induced by the free-streaming of the decay products.

The primary effect is the suppression of $P_m(k)$ on scales smaller than the comoving free-streaming length $\lambda_{\rm fs}$, i.e.\ for $k \gtrsim k_{\rm fs} \equiv 2\pi/\lambda_{\rm fs}$. This suppression propagates directly into $C_L^{\phi\phi}$, since the lensing potential probes the integrated matter distribution along the line of sight. The range of multipoles $L$ at which $C_L^{\phi\phi}$ is suppressed directly reflects the range of wavenumbers $k$ at which $P_m(k)$ is suppressed.

As shown in Fig.~\ref{fig:vary_Gamma}, increasing $\Gamma$ at fixed $v_{\rm kick}$ shifts the onset of suppression in $P_m(k)$ to larger $k$, consistent with the scaling $\lambda_{\rm fs} \propto (1+z_d)^{-1/2}$ \cite{Bencke:2026uws}: a larger decay rate implies an earlier decay, which reduces the comoving free-streaming length and pushes the suppression to smaller scales. Correspondingly, the suppression in $C_L^{\phi\phi}$ shifts to larger $L$. The amplitude of the suppression also increases with $\Gamma$, as earlier decays allow more time for the decay products to free-stream and wash out perturbations.

As shown in Fig.~\ref{fig:vary_vkick}, increasing $v_{\rm kick}$ at fixed $\Gamma$ shifts the onset of suppression to smaller $k$ (larger scales), directly reflecting the larger comoving free-streaming length. At $v_{\rm kick} = 10^{-5}c$, the suppression is negligible across the entire range of $k$ shown, consistent with the $\Lambda$CDM limit. As $v_{\rm kick}$ increases, the suppression deepens and its onset shifts to progressively larger scales, with the corresponding suppression in $C_L^{\phi\phi}$ shifting to smaller $L$. Together, these results demonstrate that \texttt{CLASSIER-DDM} accurately captures the key signatures of DDM across a wide range of model parameters.

\section{Discussion}
\label{sec:discussion}

We have presented \texttt{CLASSIER-DDM}, an extension of the Boltzmann solver \texttt{CLASSIER} that implements the decaying dark matter model via the integral-equation approach of Refs.~\cite{Kamionkowski:2021njk, Ji:2022iji, Lee:2025zym}. The code handles generic two-body decays with arbitrary masses of the decay products, naturally encompassing both massless (dark radiation) and massive (warm) decay products. We have verified that \texttt{CLASSIER-DDM} recovers the dark radiation results of \texttt{CLASS} in the massless limit, providing a robust cross-check of the implementation. \texttt{CLASSIER-DDM} is thus a general and versatile tool for studying the cosmological signatures of decaying dark matter across a wide range of model parameters.

The integral-equation approach developed for generic non-cold relics in Ref.~\cite{Lee:2025zym} is here extended to the DDM model. The extension requires careful treatment of the momentum-dependent decay time $\tau_q$, which sets a different onset of perturbations for each momentum $q$. The convergence tests presented in Sec.~\ref{sec:performance} show that $N_{\rm iter}=2$--$3$ iterations are sufficient to achieve $\mathcal{O}(0.1\%)$ convergence for the parameter space of interest, with more iterations needed only for the extreme parameter combinations, and $\mathcal{O}(1\,{\rm min})$ runtimes per evaluation making parameter estimation with the DDM model numerically tractable.

The physical results presented in Sec.~\ref{sec:results} illustrate the key signatures of the DDM model on cosmological observables. The primary effect is the suppression of power on scales smaller than the comoving free-streaming length $\lambda_{\rm fs}$, which manifests as a gradual roll-off in $P_m(k)$ and a corresponding suppression in $C_L^{\phi\phi}$ at multipoles $L \gtrsim \ell_{\rm fs}$. The scale of the suppression is set by $\Gamma$ (through the decay redshift $z_d$) and $v_{\rm kick}$ (through the comoving free-streaming length), while the amplitude is primarily set by $f$. However, the suppression of CDM perturbations exceeds the naive expectation of $1-f$, as the free-streaming decay products escape from gravitational potential wells and reduce the growth rate of the remaining CDM component, making the total suppression larger than one would expect from matter removal alone.

The oscillatory features seen in the decay product density perturbations in the middle panel of Fig.~\ref{fig:pertevo_Gamma} are a distinctive signature of the DDM model, arising from interference between contributions from decay products with different momenta $q$, each of which is produced at a different time $\tau_q$ and free-streams a different comoving distance $\chi_q(\tau_q, \tau)$, accumulating a different phase $k\chi_q(\tau_q, \tau)$ by the time $\tau$. The superposition of these contributions leads to interference, with a characteristic period scaling as $1/(kv_{\rm kick})$. These oscillatory features may leave subtle imprints on cosmological observables at small scales, particularly when the decaying dark matter constitutes a significant fraction of the total dark matter, and we leave the investigation of their observational signatures to future work.

Several extensions of \texttt{CLASSIER-DDM} are natural. Extending the analysis to smaller scales, using probes such as the Lyman-$\alpha$ forest \cite{Wang:2013rha,Fuss:2022zyt,Zhao:2026wxi}, weak lensing \cite{DES:2017myr,LSSTDarkMatterGroup:2019mwo,DES:2021wwk,Sugiyama:2023fzm}, and strong lensing \cite{Gilman:2026uvq}, would further constrain the DDM parameter space. In particular, DDM models with free-streaming on sub-galactic scales are of interest for addressing small-scale structure anomalies \cite{Abdelqader:2008wa, Bullock:2017xww, Ando:2021fhj}, and \texttt{CLASSIER-DDM} provides the tools to accurately model such scenarios. On the theoretical side, the framework could be generalized to handle decays with more than two products, enabling the study of a broader class of DDM models. More broadly, this work demonstrates the versatility and effectiveness of the integral-equation approach for modeling non-cold relics in cosmological perturbation theory, handling non-trivial phase-space distributions without any model-specific approximations. We expect the approach to find further applications in the study of exotic dark matter models and other beyond-$\Lambda$CDM scenarios where accurate and efficient treatment of non-cold species is required.

\section*{Acknowledgements}

This work was supported at JHU by NSF Grant No.\ 2412361, NASA ATP Grant No.\ 80NSSC24K1226, and the Templeton Foundation grant no.\ 62840. NL was supported by the Horizon Fellowship from Johns Hopkins University. JLB acknowledges funding from the project UC-LIME (PID2022-140670NA-I00), financed by MCIN/AEI/ 10.13039/501100011033/FEDER, UE. The Instituto de Física de Cantabria (IFCA) has been accredited as Unidad de Excelencia María de Maeztu, a distinction awarded by the Spanish State Research Agency for the period 01/06/2026 - 31/05/2030

\section*{Data Availability}

The data that support the findings of this article are openly available \cite{classier-ddm}.

\appendix

\section{Phase-space density for more than two decay products}
\label{appendix:f0-general}

If the particle has more than two decay products, then the final-state particles will have a distribution $dN/dp$ (with $\int (dN/dp)dp=1)$ of momenta $p$ in the rest-frame of the decaying particle.  The homogeneous distribution function evolves as
\begin{equation}
     \frac{\partial f_0(q,\tau)}{\partial\tau} = \frac{T_0 \Gamma }{4\pi q^2} \frac{dN}{dp}\left( p=\frac{T_0 q}{a(\tau)} \right) N_\chi(\tau),
\end{equation}     
which follows from multiplying the right-hand side of Eq.~(\ref{eqn:uniformevolution}) by $dN/dp$ and then integrating over $dp$.  The distribution at any given $\tau$ is then
\begin{equation}
   f_0(q,\tau) = \frac{T_0 \Gamma }{4\pi q^2} \int_0^{\tau}\, d\tau'\frac{dN}{dp}\left( p=\frac{T_0 q}{a(\tau')} \right) N_\chi(\tau').
\end{equation}

\section{Direct convolution with small $\xi$'s}
\label{appendix:approx-direct-conv}

The spherical Bessel functions and their second derivatives can be approximated when $\xi$ is small,
\begin{eqnarray}
j_0(\xi) &\approx& 1-\frac{\xi^2}{6},
\qquad j''_0(\xi) \approx -\frac{1}{3} + \frac{\xi^2}{10},\nonumber\\
j_1(\xi) &\approx& \frac{\xi}{3} - \frac{\xi^3}{30},
\qquad j''_1(\xi) \approx -\frac{\xi}{5} + \frac{\xi^3}{42},\nonumber\\
j_2(\xi) &\approx& \frac{\xi^2}{15}-\frac{\xi^4}{210},
\qquad j''_2(\xi) \approx \frac{2}{15} - \frac{2\xi^2}{35}.
\end{eqnarray}
Substituting these into Eq.~\eqref{eq:I-convolution} turns the convolutions into simple integrals, resulting in
\begin{eqnarray}
 (\Delta f)^{\rm cont}_0(\xi) &\approx& \mathcal{J}_1^0(\xi) - \frac{\mathcal{J}_1^2(\xi)}{6} - \frac{\mathcal{J}_2^0(\xi)}{3} + \frac{\mathcal{J}_2^2(\xi)}{10}\nonumber\\
 (\Delta f)^{\rm cont}_1(\xi) &\approx& \frac{\mathcal{J}_1^1(\xi)}{3} - \frac{\mathcal{J}_1^3(\xi)}{30} - \frac{\mathcal{J}_2^1(\xi)}{5} + \frac{\mathcal{J}_2^3(\xi)}{42}\nonumber\\
 (\Delta f)^{\rm cont}_2(\xi) &\approx& \frac{\mathcal{J}_1^2(\xi)}{15} - \frac{\mathcal{J}_1^4(\xi)}{210} + \frac{2\mathcal{J}_2^0(\xi)}{15} - \frac{2\mathcal{J}_2^2(\xi)}{35},~~~~~~~~
\end{eqnarray}
where
\begin{eqnarray}
\mathcal{J}_m^n(\xi) \equiv \sum_{k=0}^n \binom{n}{k}(-1)^{n-k}\; \xi^k \;\mathcal{I}_m^{(n-k)}(\xi),
\end{eqnarray}
and
\begin{eqnarray}
\mathcal{I}_1^{(i)}(\xi) &\equiv& -\frac{1}{kq}\frac{d \widetilde{f}_0}{d \ln q}  \int_{0}^{\xi} \, d\xi' \, (\xi')^i\; \epsilon(\xi') \;\dot{\eta} (\xi'),\nonumber\\
\mathcal{I}_2^{(i)}(\xi) &\equiv& -\frac{1}{kq}\frac{d \widetilde{f}_0}{d \ln q}  \int_{0}^{\xi} \, d\xi' \, (\xi')^i\; \epsilon(\xi') \;\frac{\dot{h}(\xi') + 6\dot{\eta}(\xi')}{2}.\nonumber\\
\end{eqnarray}
We apply this for $\xi<1$. This approximation can be straightforwardly extended to higher order by including additional terms in the Taylor expansion of the spherical Bessel functions.

\section{Fluid approximation in Ref.~\cite{FrancoAbellan:2020xnr}}
\label{appendix:comparison}

\begin{figure}[ht!]
\centering
\includegraphics[width=\linewidth]{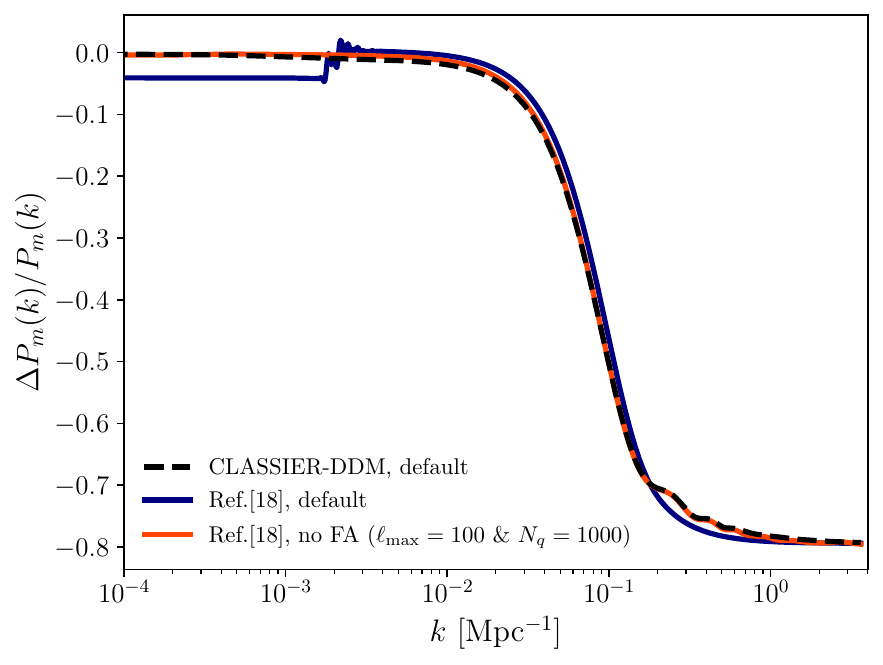}
\caption{Fractional deviation of the matter power spectrum from $\Lambda$CDM, $\Delta P_m(k)/P_m(k)$, for the DDM model with one massless and one massive decay product, with $f=1$, $\Gamma=H_0$, $\varepsilon_1=0.99$, $\varepsilon_2=0$ (corresponding to $v_{\rm kick}\approx 0.01c$), with the standard $\Lambda$CDM parameters fixed at their fiducial values, including $H_0$. We compare \texttt{CLASSIER-DDM} with default settings (black, dashed) against the code of Ref.~\cite{FrancoAbellan:2020xnr} with default settings including the fluid approximation (navy) and without the fluid approximation at high precision ($\ell_{\rm max}=100$ \& $N_q=1000$, orange). \texttt{CLASSIER-DDM} agrees well with the high-precision result of Ref.~\cite{FrancoAbellan:2020xnr}, validating our implementation in the massless decay product limit.}
\label{fig:comparison_FA}
\end{figure}

In this appendix, we illustrate the impact of the fluid approximation for decay product perturbations employed in Ref.~\cite{FrancoAbellan:2020xnr} on the predicted matter power spectrum suppression. Figure~\ref{fig:comparison_FA} shows $\Delta P_m(k)/P_m(k)$ for $f=1$, $\Gamma=H_0$, $\varepsilon_1=0.99$, $\varepsilon_2=0$ (corresponding to $v_{\rm kick}\approx 0.01c$), with the standard $\Lambda$CDM parameters fixed at their fiducial values, including $H_0$.

The result of Ref.~\cite{FrancoAbellan:2020xnr} with default settings, which employs the fluid approximation, shows percent-level deviations from the high-precision result without the fluid approximation ($\ell_{\rm max}=100$, $N_q=1000$), particularly on small scales. Notably, the fluid approximation does not capture the oscillatory features in $P_m(k)$, which arise from the free-streaming of the decay products. \texttt{CLASSIER-DDM}, which requires no such approximation, agrees well with the high-precision result.

\bibliography{mybib}

\end{document}